\documentclass[twocolumn,prl,showpacs,preprintnumbers]{revtex4}
\usepackage{graphicx}

\begin{document}

\preprint{Draft version 1}

\title{Entanglement from the Dynamics of an Ideal Bose Gas in a Lattice}
\author{Sougato Bose}
\affiliation{Department of Physics and Astronomy, University College
London, Gower St., London WC1E 6BT, UK}



\begin{abstract}
We show how the remotest sites of a finite lattice can be entangled,
with the amount of entanglement exceeding that of a singlet, solely
through the dynamics of an ideal Bose gas in a special initial state
in the lattice. When additional occupation number measurements are
made on the intermediate lattice sites, then the amount of
entanglement and the length of the lattice separating the entangled
sites can be significantly enhanced. The entanglement generated by
this dynamical procedure is found to be higher than that for the
ground state of an ideal Bose gas in the same lattice. A second
dynamical evolution is shown to verify the existence of these
entangled states, as well entangle qubits belonging to well
separated quantum registers.
\end{abstract}


\maketitle


   One of the aims in the field of quantum information is
   to set up entanglement between locations separated by some distance, and in
   general, greater the separation and more the {\em amount} of this entanglement, the better.
   While photons are best for long
   distance entanglement distribution, for short distances (such as for linking quantum registers) other alternatives
   are important \cite{wineland}. In this context, the {\em dynamics} of spin chains have been proposed for the
   distribution of entanglement over a distance of several lattice sites
(For example, Refs.\cite{bose,christandl,yung} to mention a very
few). However, as the number of possible states of a spin in a spin
chain is low, the amount of entanglement that can be dynamically
generated and distributed through a single spin chain channel in a
limited time is restricted. In this paper, as an alternative to spin
chains, we suggest the use of the dynamics of an ideal gas of $M$
bosons in a lattice to generate and distribute entanglement between
its remotest sites. Note that our dynamics, when followed by certain
occupation number measurements, will create high but ``finite"
dimensional entangled states (with entanglement $\sim
\log_2{\sqrt{M}}$+Const.) which are qualitatively very different
from the infinite dimensional Gaussian entangled states which can be
   generated dynamically through harmonic oscillator
   chains
\cite{plenio1}.

Another motivation for the current work originates from the
literature on entangling Bose-Einstein condensates (BECs) of gaseous
atoms/molecules in distinct traps (
Refs.\cite{dunningham1,dunningham3,milburn,deb,dunningham-philips}
to mention a very few) where usually small lattices or continuous
variable entanglement are considered. Can we use lattice dynamics to
create entanglement between traps separated by {\em several
intervening lattice sites}? Here we show how to accomplish this
without either the physical movement of traps or any local
modulation of the lattice parameters.

\begin{figure}
\begin{center}
\includegraphics[width=3.5in, clip]{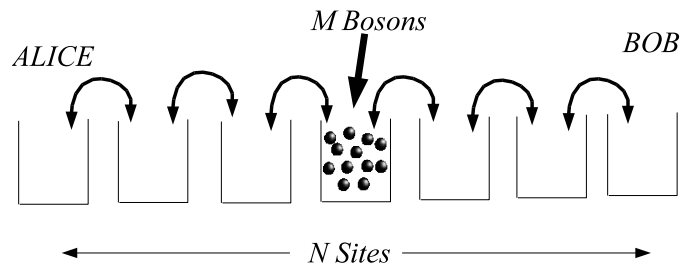}
 \caption{Our setup of creating entanglement between sites $1$ and $N$ of a 1D lattice. One simply
 starts with $M$ bosons in the $(N+1)/2$th lattice sites and allows dynamical evolution of the system to create entanglement
 between the sites $1$ and $N$. Alice and Bob who have access to the sites $1$ and $N$ can use the entanglement for quantum
 communications of for linking distinct quantum registers.}
\label{simplest-scheme}
\end{center}
\end{figure}

    We consider a one dimensional lattice of $N$ sites, where the aim is to
establish a significant amount of entanglement between sites $1$
and $N$. We choose $N$ to be odd and initially place $M$ bosons in
the $\frac{N+1}{2}$th site of the lattice and keep all the other
sites empty. Physically this corresponds to a Fock state
$|M\rangle$ on the $\frac{N+1}{2}$th site of the lattice and a
vacuum state in all the other sites. In terms of boson creation
operators $a_j^{\dagger}$ which create a boson in the $j$th
lattice site the initial state is thus
\begin{equation}
|\Psi(0)\rangle=\frac{(a_{\frac{N+1}{2}}^{\dagger})^M}{\sqrt{M!}}|0\rangle.
\end{equation}
This special initial state greatly simplifies the calculations and
can be regarded as a generalization of a single spin flip at the
midpoint of a spin chain, which has been studied for entanglement
generation \cite{yung}. We assume the bosons to be essentially
non-interacting during the time-scale of our scheme {\em i.e.,}
their collection is an {\em ideal Bose gas}. Then the Hamiltonian
of the system in the lattice is
\begin{equation}
H=J\sum_{j=1}^N (a^{\dagger}_j a_{j+1}+a_j a^{\dagger}_{j+1}).
\end{equation}
The state of each boson then evolves independently in the lattice
({\em i.e.,} each boson evolves as if it was hopping alone in an
otherwise empty lattice) as $a_{\frac{N+1}{2}}^{\dagger}\rightarrow
\sum_{j=1}^N f_j(t) a_j^{\dagger}$ where $f_j(t)$ is the amplitude
of the transfer of a single boson from the $\frac{N+1}{2}$th site to
the $j$th site in time $t$. Thus the state of the $M$ boson system
at time $t$ is
\begin{equation}
|\Psi(t)\rangle=\frac{(\sum_{j=1}^N f_j(t)
a_j^{\dagger})^M}{\sqrt{M!}}|0\rangle. \label{MBosonState}
\end{equation}
The evolution amplitudes $f_j(t)$ are identical to that of a XY
spin chain in the single excitation sector, and is given
\cite{christandl,yung-leung-bose} by
\begin{equation}
f_j(t)=\frac{2}{N+1}\sum_{k=1}^N\{\sin{\frac{\pi
k}{2}}\sin{\frac{\pi kj}{N+1}}\}e^{i2Jt\cos{\frac{k\pi}{N+1}}}.
\label{fj}
\end{equation}
Eqs.(\ref{MBosonState}) and (\ref{fj}) give the complete time
evolution of the $M$ boson state analytically.

  Our task is now to calculate how much entanglement exists between sites $1$ and $N$
in the state of Eq.(\ref{MBosonState}) as a function of time and
find a time at which this entanglement is large. We will thus have
to calculate the reduced density matrix of the states of sites $1$
and $N$ by tracing out the state of all the other sites. To
accomplish that we adopt a strategy from Ref.\cite{simon}, which
evaluates the entanglement between two regions of a Bose-Einstein
condensate in its ground state in a {\em single} trap. The strategy
is to define new creation/annihilation operators by combining the
operators $a^{\dagger}_j$ as
$E^{\dagger}=f_1(t)(a_1^{\dagger}+a_N^{\dagger})/\sqrt{2|f_1(t)|^2}$
and $L^{\dagger}=\sum_{j=2}^{N-1} f_j(t)
a_j^{\dagger}/\sqrt{1-2|f_1(t)|^2}$, which are valid bosonic
creation operators satisfying
$[E,E^{\dagger}]=1,~[L,L^{\dagger}]=1$. Noting that the symmetry of
our problem implies $f_1(t)=f_N(t)$, we expand the expression of
$|\Psi(t)\rangle$ in Eq.(\ref{MBosonState}) in terms of the above
operators to get
\begin{eqnarray}
|\Psi(t)\rangle &=&
\sum_{r=0}^M\sqrt{^MC_r}(\sqrt{2}|f_1(t)|)^r(\sqrt{1-2|f_1(t)|^2})^{M-r}\nonumber\\
&\times&  |\psi_r\rangle_{1N}|\phi_r\rangle_{2...N},
\end{eqnarray}
where we have substituted
$|\psi_r\rangle_{1N}=\frac{(E^{\dagger})^r}{\sqrt{r!}}|0\rangle$ and
$|\phi_r\rangle_{2...N}=\frac{(L^{\dagger})^{M-r}}{\sqrt{(M-r)!}}|0\rangle$.
Noting that the set of states $\{|\phi_r\rangle_{2...N}\}$
represents an orthonormal set in the space of states of the sites
$2$ to $N-1$, we have the the reduced density matrix of sites $1$
and $N$ to be
\begin{equation}
\rho(t)_{1N}=\sum_{r=0}^MP_r(t) |\psi_r\rangle\langle \psi_r|_{1N},
\end{equation}
where $P_r(t)=~^MC_r(2|f_1(t)|^2)^r(1-2|f_1(t)|^2)^{M-r}$. We can
write $|\psi_r\rangle_{1N}$ in terms of the occupation numbers as
\begin{equation}
|\psi_r\rangle_{1N}=\frac{1}{2^{r/2}}\sum_{k=0}^r\sqrt{^rC_k}|k\rangle_1|r-k\rangle_N.
\end{equation}
Note that the only time dependence of the state $\rho(t)_{1N}$ stems
from $f_1(t)$, maximizing which over a long period of time is a
pretty good strategy for obtaining a time $t_h$ such that
$\rho(t_h)_{1N}$ is highly entangled. The maximization ensures that
the proportion of the state $|\psi_M\rangle_{1N}$, which has the
most entanglement among the set of states $\{|\psi_r\rangle_{1N}\}$,
is the highest possible in $\rho(t_h)_{1N}$. Ideally, the lattice
dynamics should be frozen at $t_h$, say by globally raising the
barriers between all wells of the lattice, so that Alice and Bob can
utilize $\rho(t_{h})_{1N}$ for quantum communications or linking
quantum registers. For the smallest non-trivial lattice ($N=3$) we
know that $f_1(t)=1/\sqrt{2}$ at $t=\pi/2J\sqrt{2}$ from the XY
model \cite{yung-leung-bose}. For this case, the state
$|\psi_M\rangle_{13}$ is generated between sites $1$ and $3$ at
$t=\pi/2J\sqrt{2}$ whose entanglement can be made to grow without
limits by increasing $M$. For this special case, the advantage over
spin-1/2 chains is most evident, where only the case of $M=1$ can be
realized (with a single flip in the middle of a chain of three
spin-1/2 systems).

\begin{figure}
\begin{center}
\includegraphics[width=3.5in, clip]{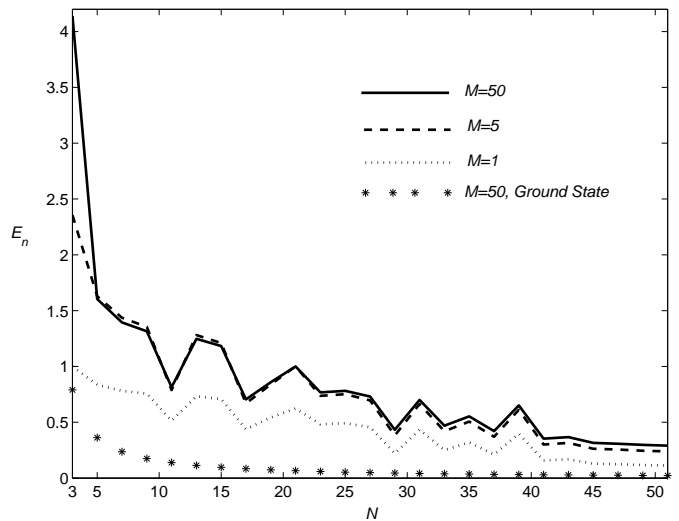}
 \caption{Entanglement (log negativity) of sites $1$ and $N$ from dynamics at an optimal time for various $N$ and $M$, and for the
ground state.}
\label{negplot}
\end{center}
\end{figure}

In general (for $N>3$), the state $\rho(t)_{1N}$ is a mixed state of
a $(M+1)\times (M+1)$ dimensional system, and the only readily
computable measure of its entanglement is the logarithmic negativity
$E_n$ \cite{vidal}, which bounds the amount of pure state
entanglement extractable by local actions from the state
$\rho(t)_{1N}$. It is the standard measure used when high
dimensional mixed entangled states arise \cite{plenio1}. The $E_n$
of $\rho(t)_{1N}$ for an appropriately chosen time $t=t_{h}$ (see
above) are plotted in Fig.\ref{negplot} as a function of $N$ for
different values of $M$. This figure clearly shows that for $N$ as
high as $21$, one can generate more entanglement than that of a
singlet ($E_n=1$ for a singlet). For such modest lengths, thus one
generates more entanglement between the ends of a lattice by using a
$M>1$ boson gas than ever possible with a spin-1/2 chain, which is
the M=1 case, also plotted in Fig.\ref{negplot}. The advantages of
increasing $M$ diminish, though, as one increases $N$. For $N=51$,
we see that though there is some advantage of high boson number
($M=50$) over the spin chain ($M=1$) case, this advantage does not
increase by increasing $M$ (for example, the $M=5$ and $50$ plots
are nearly coincident).

\begin{figure}
\begin{center}
\includegraphics[width=3.5in, clip]{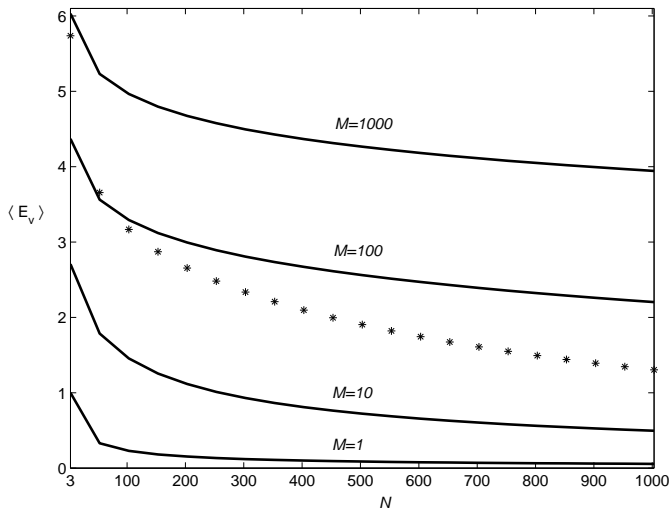}
 \caption{Average entanglement (von Neumann entropy) of sites $1$ and $N$ from dynamics followed by measurements on the intermediate lattice sites for
various $N$ and $M$. The plot with asterisks denotes the same for
the ground state for $M=1000$.}
\label{entplot}
\end{center}
\end{figure}

      Next, we slightly modify our scheme and after the dynamical evolution till
$t_{h}$, we measure the total number of bosons in the intermediate
sites (sites $2$ to $N-1$). With probability $P_r$, we will find
this number to be $M-r$, and when we do so, we will generate the
state $|\psi_r\rangle_{1N}$ between sites $1$ and $N$. Note that
$|\psi_r\rangle_{1N}$ is created between sites $1$ and $N$ whenever
a ``total" of $M-r$ bosons is found in the remaining sites {\em
irrespective} of the distribution of these $M-r$ bosons among the
sites. The amount of entanglement in the pure state
$|\psi_r\rangle_{1N}$ is given by its von Neumann entropy of
entanglement $E_v(|\psi_r\rangle_{1N})=-Tr(\rho_1\log_2\rho_1)$,
where $\rho_1=Tr_N(|\psi_r\rangle\langle \psi_r|_{1N})$, which {\em
equals} the quantity of entanglement that can be obtained as
singlets (the most useful form, say for their use as a resource for
perfect teleportation of qubits from Alice to Bob) from the state by
local actions and classical communications alone. Thus the average
von Neumann entropy of entanglement $\langle E_v \rangle=\sum_r P_r
E_v(|\psi_r\rangle_{1N})$ over all possible measurement outcomes is
operationally the most useful measure of the entanglement between
sites $1$ and $N$ in our modified scheme. This quantity has been
plotted in Fig.\ref{entplot} for various $M$ for large lengths of
lattice up to $N=1001$. We find that for $M=1000$, entanglement
nearly equal to that of $4$ singlets is generated across a distance
of $1001$ lattice sites, and this is more than $70$ times the amount
possible with a spin-1/2 chain ($M=1$) of same length. For high $N$
and $M$, we can represent $P_r$ by a Poisson distribution and know
that $f_1(t)\sim 1.7/N^{1/3}$ at $t\sim (N+0.81 N^{1/3})/4J$
\cite{bose,yung-leung-bose}. Using these, we have the analytic
expression $\langle E_v \rangle \sim \log_2\{{1.7\sqrt{M\pi
e}/N^{1/3}}\}$, whose fit with data gets better as we proceed from
the $M=10$ to the $M=1000$ plot at high $N$. For every order of
magnitude increase of $M$ we thus expect to gain an entanglement
equalling that of $\log_2{\sqrt{10}}=1.66$ singlets, which is
confirmed by the differences in $\langle E_v \rangle$ between the
$M=100$ and $1000$ plots at high $N$ (Fig.\ref{entplot}). So for
obtaining entanglement exceeding that of $10$ singlets across a
distance of $\sim 1000$ lattice sites we require $M\sim 10^7$.

    Now we proceed to compare the amount of entanglement between
sites $1$ and $N$ generated by the above schemes with that
obtainable from the ground state of an ideal Bose gas in the same
lattice (to check whether we have gained from dynamics). Ground
state entanglement between distinct regions of a Bose-Einstein
condensate in a {\em single} trap has already been investigated
\cite{simon,vedral}. In our lattice setting, the ground state of $H$
is simply $(\frac{1}{\sqrt{N}}\sum_{j=1}^N
a_j^{\dagger})^M|0\rangle/\sqrt{M!}$. It thus suffices to replace
$f_1(t)$ in the expression of $P_r(t)$ with $1/\sqrt{N}$ to change
from the dynamical state to the ground state. $E_n$ between sites
$1$ and $N$ for the ground state is plotted using asterisks in
Fig.\ref{negplot} as a function of $N$ for $M=50$. We find that the
entanglement of sites $1$ and $N$ is nearly vanishing for $N\geq
25$, and thus there is a marked advantage of using dynamics as
opposed to the ground state. One may also however, measure the
occupation numbers of sites $2$ to $N$ of the ground state to obtain
an average entanglement $\langle E_v \rangle\sim \log_2{\sqrt{M\pi
e/N}}$ between sites $1$ and $N$ (for high $M$ and $N$). Even this
is lower than that of our second scheme by $\log_2{\{1.7N^{1/6}\}}$.

  We now proceed to discuss a method to verify the pure entangled
  states $|\psi_r\rangle_{1N}$ created
  through our second scheme. One can easily verify the number correlations (the fact that if site $1$ has $k$ bosons, then
  the site $N$ has $r-k$ bosons) in $|\psi_r\rangle_{1N}$ by
  occupation number measurements. Then the only task remaining is to verify the coherence between
  the terms $|k\rangle_1|r-k\rangle_N$ in $|\psi_r\rangle_{1N}$.
  If we apply a Hamiltonian $H_{v}=\kappa a_1^\dagger a_1$ for a
  fixed time $t_{v}$ to site $1$ then the above coherence leads
  to $|\psi_r\rangle_{1N}$ becoming
  $|\psi_r^{'}\rangle_{1N}=(e^{i\theta}a_1^{\dagger}+a_N^{\dagger})^r|0\rangle/\sqrt{2^{r}r!}$
  where $\theta=\kappa t_{v}$. The ideal Bose gas is then
  allowed to evolve again in the lattice (assuming the lattice
  sites $2$ to $N-1$ have been emptied due to or after the earlier
  occupation number measurements), which results in its state
  becoming $|\Psi^{'}(t)\rangle_{1..N}=(\sum_{j=1}^{N} \{e^{i\theta}g_{1j}(t)+g_{Nj}(t)\} a_j^{\dagger})^r|0\rangle/\sqrt{2^{r}r!}$
  where $g_{lj}$ are amplitudes for a boson going from the $l$th site to the $j$th
  site. From symmetry and the formulae for evolution in XY chain,
  we have $g_{1\frac{N+1}{2}}(t)=g_{N\frac{N+1}{2}}(t)=f_1(t)$.
  Thus the probability of finding the site $\frac{N+1}{2}$ occupied can be varied from $1-(1-2|f_1(t)|^2)^{r}$
  for $\theta=0$, to $0$ for $\theta=\pi$ (as the term $a_{\frac{N+1}{2}}^{\dagger}$
   in $|\Psi^{'}(t)\rangle_{1..N}$ vanishes for $\theta=\pi$). This variation with $\theta$, whose range increases with increasing $r$ and
which can be further increased by maximizing $|f_1(t)|$,  enables us
to verify the
   coherence between the terms $|k\rangle_1|r-k\rangle_N$
  in $|\psi_r\rangle_{1N}$. One way to verify the mixed state
  $\rho(t)_{1N}$ produced by our first scheme will be to go
  through our second scheme and check that pure states
  $|\psi_r\rangle_{1N}$ are produced with probabilities $P_r$.

  Next, we provide an example of linking qubits $A$ and $B$ of distant quantum
  registers by making them interact with sites $1$ and $N$ respectively for a time $t_q$, when these sites are in the state
$|\psi_r\rangle_{1N}$. The initial state of the qubits is
$(|0\rangle_A+|1\rangle_A)\otimes(|0\rangle_B+|1\rangle_B)/2$ and
the interaction Hamiltonian is $H_{q}=g(\sigma_z^A a_1^\dagger
a_1+\sigma_z^B a_N^\dagger
  a_N)$ in which
  $\sigma_z^A$ and $\sigma_z^B$ are Pauli operators of the qubits.
  For $gt_q=\pi$, the state of the system is $\{(|0\rangle_A
  |0\rangle_B+(-1)^r|1\rangle_A
  |1\rangle_B)\otimes(a_1^{\dagger}+a_N^{\dagger})^r|0\rangle+(|0\rangle_A
  |1\rangle_B+(-1)^r|1\rangle_A
  |0\rangle_B)\otimes(a_1^{\dagger}-a_N^{\dagger})^r|0\rangle\}/\sqrt{2^{r+2}r!}$.
  As in the previous paragraph, now the state of the ideal Bose
  gas is again allowed to evolve in the lattice and
  after some time the presence of any boson the $(N+1)/2$th site is measured. If
  this site is found occupied, then, remembering the
  logic of the last paragraph, the gas must have been in the state
  $(a_1^{\dagger}+a_N^{\dagger})^r|0\rangle/\sqrt{2^{r}r!}$, which
  projects the qubits to the maximally entangled state $|0\rangle_A
  |0\rangle_B+(-1)^r|1\rangle_A
  |1\rangle_B$. The probability for this is $1/2$.  Otherwise (also with probability $1/2$), a state whose fidelity with $|0\rangle_A
  |1\rangle_B+(-1)^r|1\rangle_A
  |0\rangle_B$ is $1-(1-2|f_1(t)|^2)^{r}$ is obtained (this fidelity can be maximized by maximizing $|f_1(t)|$,
and gets better with increasing $r$). This scheme for
  entangling qubits of distinct registers is just an example. The values of $\langle E_v \rangle$ found earlier imply
  that in {\em
  principle} one should be able to extract many more singlets
  from
  $|\psi_r\rangle_{1N}$ by local actions alone.

BECs of dilute atomic gases in optical lattices form a test ground
for our protocols. Fock states can be prepared by Mott transitions
\cite{mott}, the interactions needed for that can be switched off by
a Feshbach resonance \cite{feshbach} to obtain an ideal Bose gas
(otherwise, one simply has to go between regimes were the on-site
repulsion $U>>J/M^2$ to $U<<J/M^2N$ by the global modulation of
lattice potentials). Accurate measurement of the total number of
bosons in sites $2$ to $N-1$ is possible either by using metastable
atoms \cite{meta} or potentially through a second Mott transition
involving these sites only (note that individual site occupation
numbers are not required). For the verification part, only whether
the $(N+1)/2$th site is occupied or not need to be ascertained, and
atomic fluorescence in external fields can potentially be used. By
placing all atoms in a known magnetic state when required (say by a
laser), each site can be imparted a magnetic moment proportional to
its occupation number. This, and a local magnetic field at site $1$
can be used to realize $H_v$. This magnetic moment also enables
magnetic moment based register qubits (atomic or solid state) to
interact with sites $1$ and $N$ through $H_q$. If imparting the
atoms with a magnetic moment automatically make $U>>J/M^2$ then we
have to ensure that $Ut_v$ and $Ut_q$ are integral multiples of
$2\pi$. An alternative implementation is nano-oscillators arrays
\cite{plenio1} when resonantly coupled to each other. They can be
coupled to Cooper-pair box qubits to both create and measure Fock
states \cite{irish}, and to Josephson qubits through a
Jaynes-Cummings model \cite{cleland} whose off resonant limit can
implement $H_q$. Coupled cavities in photonic crystals, where Fock
state preparation and measurements could be performed with dopant
atoms is another potential implementation \cite{angelakis}.

  I thank EPSRC for an Advanced Research Fellowship and for support through the grant
GR/S62796/01 and the QIPIRC (GR/S82176/01).



\begin{thebibliography}{99}
\bibitem{wineland}
D. Kielpinski, C. Monroe and D. J. Wineland, Nature {\bf 417}, 709
(2002).
\bibitem{bose}
S.~Bose, Phys. Rev. Lett \textbf{91}, 207901(2003); D.~Burgarth and
S.~Bose, Phys. Rev. A {\bf 71}, 052315 (2005); V. Giovannetti and D.
Burgarth, Phys. Rev. Lett. {\bf 96}, 030501 (2006); T. Boness, S.
Bose, and T. S. Monteiro, Phys. Rev. Lett. {\bf 96}, 187201 (2006);
J. Fitzsimons and J. Twamley, Phys. Rev. Lett. {\bf 97}, 090502
(2006).


\bibitem{christandl}
M.~Christandl, N. Datta, A. Ekert and A. J. Landahl, Phys. Rev.
Lett. \textbf{92}, 187902 (2004).

\bibitem{yung} M.-H. Yung and S. Bose, Phys. Rev. A {\bf 71}, 032310
(2005).


\bibitem{plenio1}
J. Eisert, M. B. Plenio, S. Bose and J. Hartley, Phys. Rev. Lett.
{\bf 93}, 190402 (2004); M.B. Plenio, J. Hartley and J. Eisert, New
J. Phys. {\bf 6}, 36 (2004).




\bibitem{dunningham1}
J. A. Dunningham and K. Burnett, Phys. Rev. A {\bf 61}, 065601
(2000); J. A. Dunningham, K. Burnett and M. Edwards, Phys. Rev. A
{\bf 64}, 015601 (2001).

\bibitem{dunningham3}
J. A. Dunningham and K. Burnett, Phys. Rev. A {\bf 70}, 033601
(2004).

\bibitem{milburn}
A. P. Hines, R. H. McKenzie and G. J. Milburn, Phys.Rev.A {\bf 67},
013609 (2003).


\bibitem{deb}
B. Deb an G. S. Agarwal, Phys. Rev. A {\bf 67}, 023603 (2003).
\bibitem{dunningham-philips}
J. A. Dunningham, K. Burnett, R. Roth and W. D. Phillips,
quant-ph/0608242.
\bibitem{yung-leung-bose}
M.-H. Yung, D. W. Leung, and S. Bose, Quantum Inf. Comput. {\bf 4},
174 (2003).

\bibitem{vidal}
G. Vidal and R. F. Werner, Phys. Rev. A {\bf 65}, 032314 (2002); M.
B. Plenio, Phys. Rev. Lett. {\bf 95}, 090503 (2005).


\bibitem{simon}
C. Simon, Phys. Rev. A {\bf 66}, 052323 (2002).


\bibitem{vedral}
L. Heaney, J. Anders and V. Vedral, quant-ph/0607069; V. Vedral and
D. Kaszilikowski, quant-ph/0606238; D. Kaszilikowski {\em et. al.},
quant-ph/0601089.

\bibitem{mott}
M. Greiner {\em et. al.}, Nature {\bf 415}, 39 (2002); C. Orzel {\em
et. al.}, Science {\bf 291}, 2386 (2001).

\bibitem{feshbach}
S. Inouye {\em et. al.} Nature {\bf 392}, 151 (1998); J. L. Roberts
{\em et. al.}, Phys. Rev. Lett. {\bf 86}, 4211 (2001).

\bibitem{meta}
A. Robert {\em et. al}, Science {\bf 292}, 461 (2001).

\bibitem{irish}
E. K. Irish and K. Schwab, Phys. Rev. B {\bf 68}, 155311 (2003).

\bibitem{cleland}
M. R. Geller and A. N. Cleland, Phys. Rev. A {\bf 71}, 032311
(2005).

\bibitem{angelakis}
D. G. Angelakis, M. F. Santos and S. Bose, quant-ph/0606159; M. J.
Hartmann, F. G. S. L. Brand$\tilde{a}$o and M. B. Plenio,
quant-ph/0606097; A. D. Greentree {\em et. al.}, cond-mat/0609050.



\end{thebibliography}
\end{document}